\documentclass[%
 aip,
 amsmath,amssymb,
 reprint,
]{revtex4-1}

\usepackage{graphicx}
\usepackage{dcolumn}
\usepackage{bm}
\usepackage{amsmath}
\usepackage[english]{babel}
\usepackage{amsfonts} 
\usepackage{xcolor}
\usepackage{dsfont}
\usepackage[normalem]{ulem}
\usepackage[utf8]{inputenc}
\usepackage[T1]{fontenc}
\usepackage{mathptmx}
\usepackage{etoolbox}

\makeatletter
\def\@email#1#2{%
 \endgroup
 \patchcmd{\titleblock@produce}
  {\frontmatter@RRAPformat}
  {\frontmatter@RRAPformat{\produce@RRAP{*#1\href{mailto:#2}{#2}}}\frontmatter@RRAPformat}
  {}{}
}%
\makeatother
\begin{document}

\preprint{AIP/123-QED}

\title[Three Carleman routes to the quantum simulation of classical fluids]{Three Carleman routes to the quantum simulation of classical fluids}
\author{S. Succi, C. Sanavio}
\affiliation{Fondazione Istituto Italiano di Tecnologia\\
Center for Life Nano-Neuroscience at la Sapienza\\
Viale Regina Elena 291, 00161 Roma, Italy}
 \email{claudio.sanavio@iit.it}
\author{R. Scatamacchia}%
\affiliation{Centro Nazionale di Meteorologia e Climatologia Aerospaziale\\
Via Pratica di Mare 45, 00040 Pomezia, Italy
}%
\author{C. de Falco}
\affiliation{MOX – Modelling and Scientific Computing, Department of Mathematics, Politecnico di Milano, Milan, Italy
}%

\date{\today}

\begin{abstract}
We discuss the Carleman approach to the quantum simulation of classical fluids,
as applied to i) Lattice Boltzmann (CLB), ii) Navier-Stokes (CNS) 
and iii) Grad (CG) formulations of fluid dynamics.
CLB shows excellent convergence properties, but it is plagued by 
nonlocality which results in an exponential depth of the corresponding 
circuit with the number of Carleman variables.  
The CNS offers a dramatic reduction of the number Carleman variables, which
might lead to a viable depth, provided locality can be preserved and 
convergence can be achieved with a moderate number of iterates also 
at sizeable Reynolds numbers. 
Finally it is argued that CG might combine the best of CLB and CNS.
\end{abstract}

\maketitle

\section{Introduction}

One of the authors (SS) had the good fortune to be in friendly relations 
with Sreeni for over three decades, a great scientific honor and a human privilege alike.
Yet, we never coauthored any paper until last year and on a topic which is no mainstream
to either of us: quantum computing, and more specifically, quantum computing for fluids.  
Hence, on the occasion of his Festschrift, the same author
resolved to write about this fascinating and over-challenging topic.

Quantum computing (QC) offers tantalizing prospects for cracking problems which lie far beyond
reach of any foreseeable classical computer, typical examples in point being
quantum many-body sytems as they occur in quantum material science and quantum chemistry 
\cite{QC,Preskill2018,FEYN}.

The main point rests with the very peculiar property of quantum system to live in a 
linear superposition of states, each of which carries independent information, along
with the capability of processing it, independently. 
In a nutshell: built-in quantum parallelism.
The potential is huge: the Hilbert space of a N-bit quantum system contains
$2^N$ quantum states, which can be stored and processed in the form of 
a string of $N$ qubits, i.e units of quantum information which, unlike 
classical bits, can take {\it any} value between $0$ and $1$.
As a result, quantum computing offers exponential advantage over classical computing.

If this looks too good to be true, it is because it {\it is} indeed too good to 
be true: many obstacles stand on the way of the aforementioned blue-sky quantum computing scenario.
First, decoherence: in order to realize the above potential qubits need to be entangled, meaning
that any action on qubit A (Alice) is going to affect qubit B (Bob) as well, and vice-versa. 
Entanglement is a subtle form of correlation which proves very fragile against external perturbations,
the environment, the result being that entanglement decays very fast, currently in a time lapse
of hundreds of microseconds. At a processing speed of, say, one quantum update per nanoseconds,
this means a few hundred thousands operations, before the qubit "dies out".
The second technological problem is quantum noise: even if the qubit is "alive", this does
not mean that it computes error-free. The same is true for classical bits, the difference being
that the error rates is some fifteen orders of magnitude larger, something like $10^{-3}$
against the $10^{-18}$ rate of classical computers!
Recovering such gigantic gap on mere technological grounds appears rather desperate, but
error correction/mitigation algorithm may offer a robust helping hand to this purpose.
This is why, despite the daunting barriers above, research on quantum computing 
is still burgeoning on both hardware and software
fronts, over thirty years after the basic idea was first brought up \cite{Deutsch1985}.

As mentioned above, the best candidates for quantum computing are quantum many-body
systems (we focus on natural science, leaving aside paramount applications such as 
cryptography). Yet, it is only natural to wonder whether QC may contribute to solve
also hard problems in classical physics, turbulence being a prominent example in point~\cite{bharadwaj_hybrid_2023,ingelmann_two_2023,succi_ensemble_2024}
(general gravitation would come next).
That's exactly the topic where Sreeni and SS finally managed to publish our first paper
\cite{QC_POF} after three decades of unpublished scientific exchanges! 

\section{Quantum computing for fluids}

It is often heard that turbulent flows raise a ceaseless demand for
increasingly more powerful computers and improved computational methods.
The main culprit being the fact that the number of active degrees of freedom
scales like the cube of the Reynolds number, and since Reynolds number in Nature
easily exceeds millions (automobiles), billions (regional weather forecast)
and trillions (astrophysics), no foreseeable classical computer can meet this demand.
Quantum computing has potential to put this quest at rest \cite{QC_Sreeni}. 

Indeed, the  blue-sky scenario potential of quantum computing for fluids is mind-boggling:
given that a turbulent fluid at Reynolds $Re$ features $Re^3$ dynamic degrees of freedom,
the number of qubits required to represent it is given by \cite{RENE2019,QC_EPL}
\begin{equation}
q = 3 Log_2 \; Re \sim  10 Log_{10} Re 
\end{equation}
This means that a full airplane simulation ($Re \sim 10^8$) takes just $q \sim 80$
qubits, well within the {\it nominal} capabilities of current quantum hardware \cite{QIBM}.
At the top of the line (as of 2024), $q \sim 500$, would formally enable simulations
at $Re \sim 10^{500}$ which is far beyond any conceivable need in the physics of fluids!  

Again, too good to be true: the $500$ nominal (physical) qubits must be mapped back to the
actual number of {\it effective (logical)} qubits, whose ratio is often 
estimated at about $1:1000$. Even taking a more optimistic $1:100$ ratio, the Exascale
bar, which we place conventionally around $Re \sim 10^8$, would require 
$8000$ physical qubits, about an order of magnitude beyond the 
current quantum hardware capabilities. 

In the previous section we have mentioned the two major technological stumbling blocks
for quantum {\it computers}: decoherence and quantum noise.
When we move from computers to computing, other major limitations take stage.
The first is that not every problem can be formulated in terms of an efficient quantum
algorithm, meaning by this an algorithm which can be (efficiently) mapped into the circuits
of a quantum computer. Such circuits consisted of a collections 
of one and two qubit gates performing {\it linear and unitary} 
operations, in full compliance with the linear and
unitary nature of quantum mechanics.

This alone signals two major elephants in the room for quantum computing of fluids
systems: quantum mechanics is linear, unitary and hamiltonian, the physics 
of fluids is generally none of the three. 

Several strategies can be conceived to turn around these problems, but the one that 
Sreeni and SS have been exploring together is the so-called Carleman
linearization (or embedding), CL for short.

\begin{figure}
    \centering
    \includegraphics[scale = 0.43]{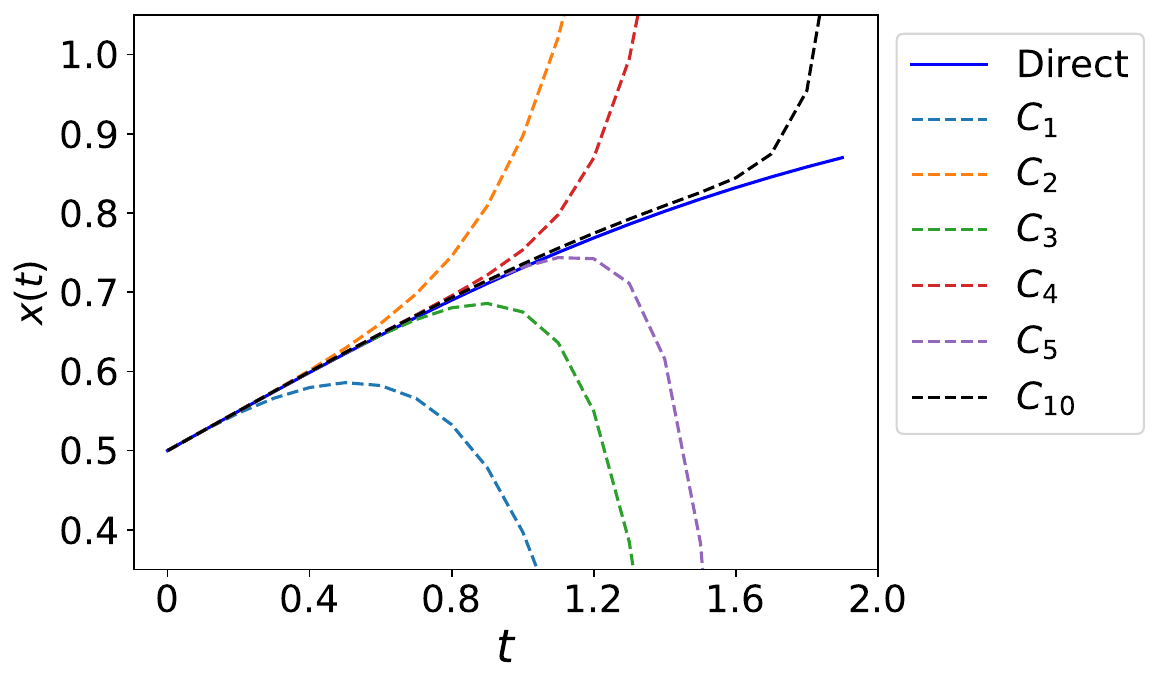}
    \caption{\label{fig:carleman_logistic} 
Convergence of the Carleman Linearization for the logistic equation at increasing
orders of the truncation. As one can see, increasing the order of truncation extends
the convergence time horizon, beyond which the solution drastically departs from
the exact one.}
\end{figure}
\section{Carleman linearization}

The general idea of the CL procedure is to turn a nonlinear finite-dimensional system
into an infinite-dimensional linear one.
Let us illustrate the point with specific reference to the simplest and
yet representative example of the logistic equation:
\begin{equation}
\label{LOGI}
\dot x = x (1-Rx)\;\;\;x(0)=x_0
\end{equation}
where $R$, the competition rate, is a measure of the nonlinearity.
The exact solution is readily derived
\begin{equation}
\label{EXA}
x(t) = \frac{x_0 e^t}{1+Rx_0 (e^t-1)} 
\end{equation}
and reaches its stable time asymptotic value $x_{\infty}=1/R$ for $t \gg 1$.
The Carleman procedure consists in renaming $x^{(1)} \equiv x$ and 
$x^{(2)} \equiv x^2$, so that the logistic equation takes the linear form 
\begin{equation}
\dot x^{(1)} = x^{(1)}-Rx^{(2)}
\end{equation}

Iterating the procedure to the $k$-th order delivers:
\begin{equation}
\dot x^{(k)} = k x^{(k)}-kRx^{(k+1)}
\end{equation}

This is an endless hierarchy, which is then truncated at a given 
order $K$ by setting $x^{K}=0$, in the hope that the truncated solution 
captures the essential behaviour of the exact one.
In practice, it can be shown that the Carleman hierarchy is just another way of
representing the exact solution as an infinite power series in the
saturating term $Rx_0(e^t-1)$. Each successive iteration prolongs the time-horizon
of convergence, see Fig.~\ref{fig:carleman_logistic}, beyond which convergence
is abruptly lost.

It should be observed that convergence is controlled by the parameter 
$Rx_0$, namely the ratio $x_0/x_{\infty}$ between the initial and time 
asymptotic values.

Despite its simplicity, the logistic equation delivers a few useful
hints for the Carleman linearization of fluids.
First, it is a quadratic non-linearity and second it shows 
that the time-horizon of the Carleman series depends on the strength 
of the nonlinearity via the ratio $x_0/x_{\infty}$.  
Third, the logistic equation bears a closed similarity
to the collision operator of kinetic equations.
This observation prompted out the earliest attempts to apply
the Carleman procedure to the Lattice formulation of fluids, as
we describe next.     

\subsection{Carleman Lattice Boltzmann}

The Carleman lattice Boltzmann (CLB) procedure has been first advocated in \cite{WAEL22}
and lately further explored by a number of authors \cite{BUDI1,BUDI2,MARGIE,SANAV}. 
The LB equation~\cite{BENZI92} takes the following form
\begin{equation}
\label{LB}
f_i(\vec{x}+\vec{c}_i,t+1) = (1-\omega) f_i + \omega f_i^{eq}
\end{equation}
where $f_i \equiv f_i(\vec x,t)$ is the probability of finding a fluid parcel ("population") at position
$\vec x$ and time $t$ with discrete velocity $\vec c_i$ and $i=0, \dots, n_v$, chosen from a suitably discrete lattice. 
In the above $f_i^{eq}$ is a local--quadratic function of the flow field
$J_a(\vec x,t)=\sum_i c_{ia} f_i(\vec x,t)/\rho$, where $\rho(\vec x,t)=\sum_i f_i(\vec x;t)$ is the local fluid density. 
In equations:
\begin{equation}
\label{LB_eq}
f_i^{eq}(\vec x,t) = w_i (\rho + J_a c_{ia} + \frac{J_a J_b}{\rho} (c_{ia}c_{ib}- \delta_{ab})) 
\end{equation}
where latin indices label spatial components and the discrete velocities are rescaled
with the sound speed $\vec c_i \rightarrow \vec c_i/c_s$. The parameter $\omega$ is inversely proportional to the relaxation time of the fluid and to the viscosity $\nu $, which can be obtained through the formula 
\begin{equation}\label{eq:viscosity_omega}
\nu  = c_s^2\bigg{(}\frac{1}{\omega}-\frac{1}{2}\bigg{)}.
\end{equation}. 

In order to deal with the nonpolynomial term $\rho^{-1}J_aJ_b$ we apply the {\it{weakly-compressible}} fluid approximation, $\rho\sim 1$, substituting $\rho^{-1}\approx2-\rho.$ This makes the equilibrium functions $f_i^{\text{eq}}$ cubic polynomials of the LB distribution functions. 
The LB method  has met with spectacular success for the simulation of fluid problems
across a broad spectrum of applications, scales and regimes \cite{MELCH2010,OUP18}.
Its main virtues are as follows: 
First, free-streaming proceeds along discrete characteristics 
$\vec \Delta x_i = \vec c_i \Delta t$
and it is floating-point free, for it amounts to shifting the populations from the
source site $\vec x$ to the corresponding lattice destination 
$\vec x_{i} = \vec x + \vec c_{i} \Delta t$.
Since streaming always lands the population on lattice sites, it 
is exact on a computer and free of most lattice artifacts.
Second, equilibria are nonlinear (quadratic) but local, the result being that dissipation
is an emergent property which does not require second order spatial derivatives.

When it comes to quantum computing, a third point pops out: the nonlinearity of the LB
method is governed by local quadratic terms $J^2$ versus the linear ones, $J$, thus
implying that the nonlinearity is formally controlled by the Mach number instead of the
Reynolds number, which makes a huge difference.

As a result, it has been argued that CLB (Carleman-Lattice Boltzmann)
might work better than CNS (Carleman-Navier Stokes) \cite{MARGIE}.  
Indeed, both \cite{MARGIE} and \cite{SANAV} report excellent convergence
up to Reynolds numbers of the order $10-100$, with just two or three Carleman
iterates. 

\begin{figure*}
\includegraphics[scale = 0.35]{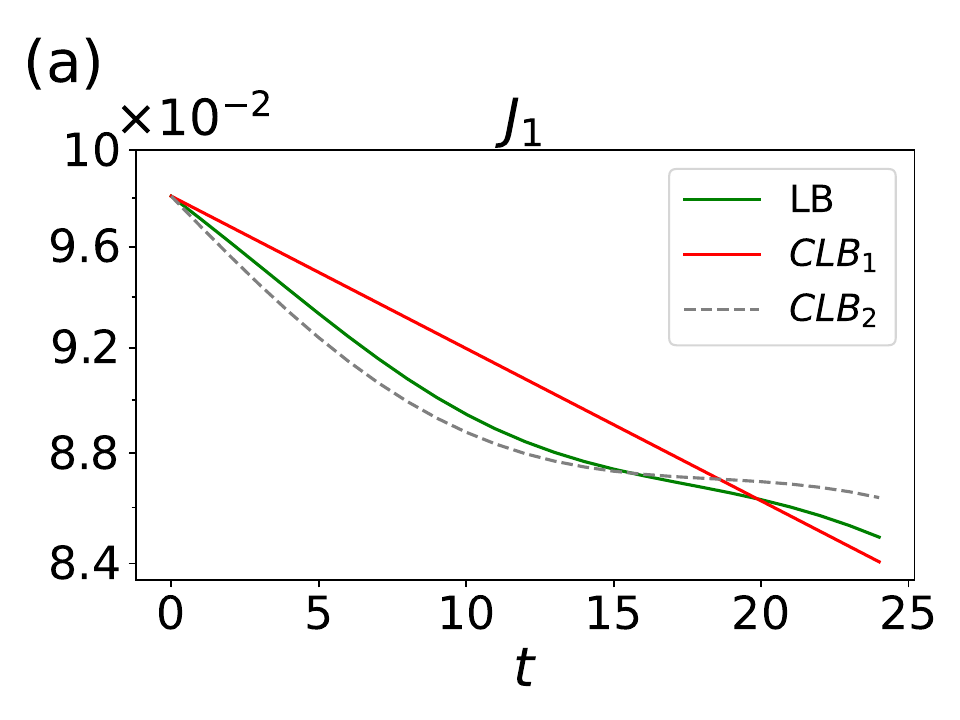}
\includegraphics[scale = 0.35]{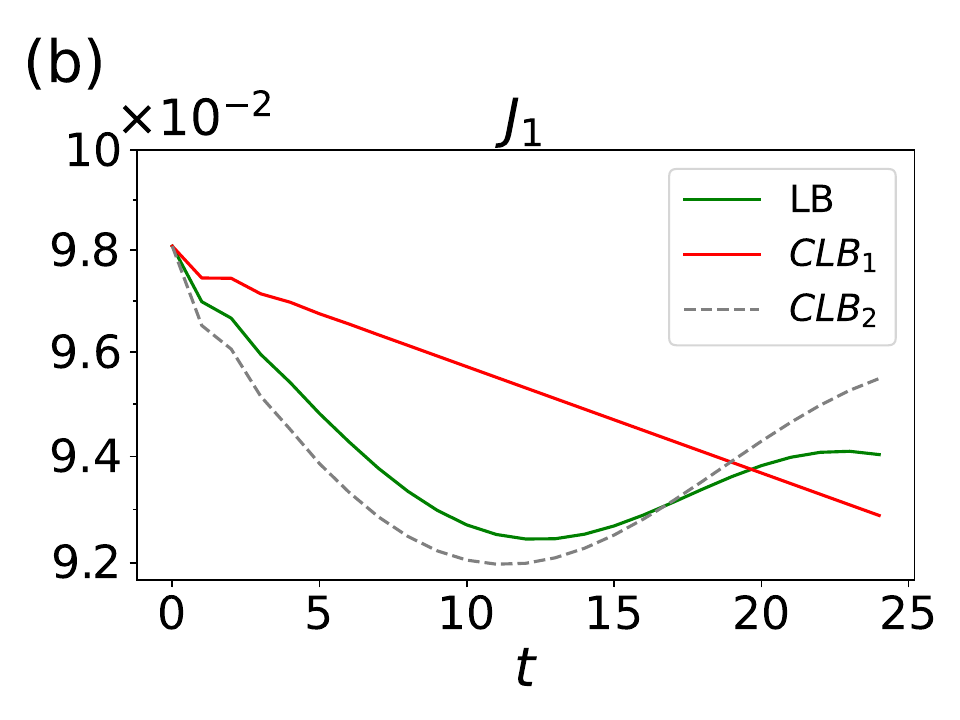}
\includegraphics[scale = 0.35]{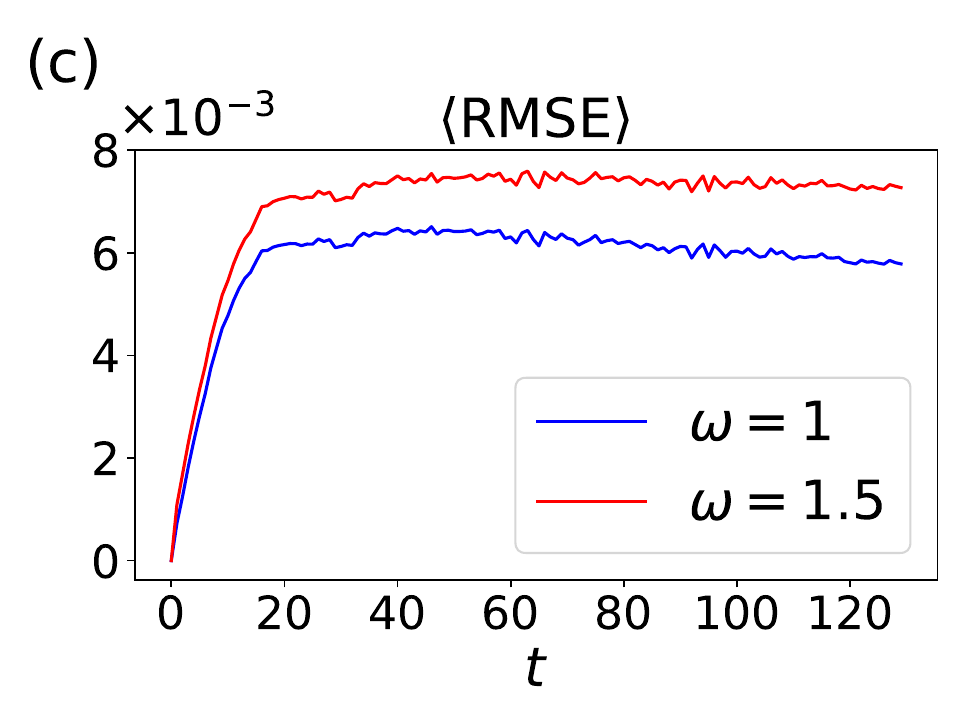}
\caption{\label{fig:CLB} The dynamics of the Kolmogorov-like flow on a $32\times32$ grid, 
plotted for a point of the grid, evolved with the LB method and comparison with 
CLB at first (CLB$_1$) and second (CLB$_2$) order for  $\omega=1$ in (a) and $\omega=1.5$ in (b). 
In (c) the average Root Mean Squared Error $\langle RMSE \rangle$ between Carleman at second order and exact lattice Boltzmann. 
The parameters of the Kolmogorov flow at $t=0$ 
are $u_1 = u_2 = 0.1$, 
$k_1=k_2=1.$} 
\end{figure*}

We tested CLB on a 2D lattice of $32\times32$ points in the grid $G$, where the initial conditions were set to

\begin{equation}\label{eq:kolmogorov-like}
\rho(\vec x) = 1,\quad J_a = u_a\cos{(k_a x_{b})}, 
\end{equation}

\noindent for each of the gridpoints $\vec x\in G$, with $a,b=1,2$ and $a\neq b$.
The values of the distribution functions $f_i$ with are obtained accordingly to standard methods.
This is shown in Figs.~\ref{fig:CLB}(a) and ~\ref{fig:CLB}(b), where we plot the macroscopic density $J_1(\tilde{x},t)$ for a random point $\tilde{x}$, found by the exact LB method and the corresponding approximate solutions provided by the Carleman linearization at orders 1 and 2. The two figures refer to different choices of the parameters $\omega$, namely $\omega=1$ in Fig.~\ref{fig:CLB}(a) and $\omega = 1.5$ in Fig.~\ref{fig:CLB}(b). 
The other parameters and the choice for the initial conditions are detailed in the caption. 
In Figure~\ref{fig:CLB}(c) we plot the average deviation  $\langle RMSE\rangle$ (Root Mean Squared Error) between the results obtained by the exact LB calculation and the ones obtained by the Carleman linearization of different orders. 
It is defined as

\begin{equation}\label{eq:deviation}
\langle RMSE \rangle=\sum_a\frac{1}{n_v}\sqrt{\sum_{x}\frac{(J_a^{\text{LB}}-J_a^{\text{CLB}})^2}{G}},\quad{a=1,2}
\end{equation}

\noindent Fig.~\ref{fig:CLB}(c) shows the results for $\omega=1$ and $\omega=1.5$ respectively. It’s worth to notice that although the Mach number should be the relevant parameter for accounting the non-linearity of the system in the LB framework, larger deviations are obtained for higher Reynolds (lower viscosity).

Unfortunately, the free-streaming also brings an undesired non-local coupling
of the Carleman variables. This is easily understood by considering for instance
the local Carleman pair $f_{ij}(\vec x,\vec x;t) \equiv f_i(\vec x,t)f_j(\vec x,t)$. 
At time $t+1$ ($\Delta t=1$), this flies into the nonlocal pair 
$f_i(\vec x+\vec c_i,t+1)f_j(\vec x+\vec c_j,t+1) \equiv f_{ij}(\vec x+\vec c_i,\vec x+\vec c_j,t+1)$.

\begin{figure}
\includegraphics[scale = 0.5]{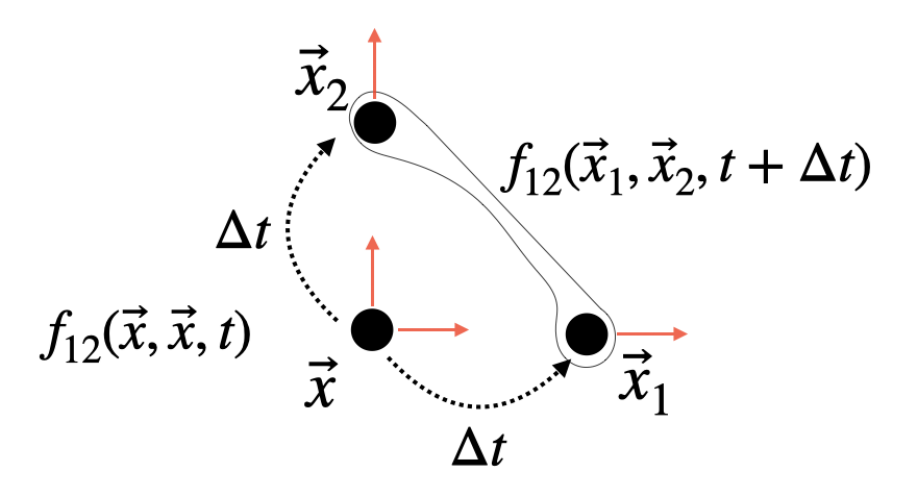}
\caption{{\label{fig:non_locality_CLB} The streaming of the second order local variables $f_{12}(\vec x,\vec x,t)$ leads to a nonlocal variables $f_{12}(\vec x_1,\vec x_2,t+\Delta t)$.}}
\end{figure}

A sketch of this feature is shown in Fig.~\ref{fig:non_locality_CLB}. 

Due to this nonlocality, the number of CLB variables in a block-time algorithm
marching from $t=0$ all the way to the end time $t$,
scales like $(n_vG)^k$, which is totally unviable despite the low-$k$ convergence. 
Ultimately this amounts to an exponential depth problem of the corresponding 
quantum circuit, which is found to lie close to the theoretical upper 
bound $De \sim 4^q = N^2$ \cite{SANAV}.
Of course, a single step algorithm from $t$ to $t+\Delta t$
would feature $(n_v)^k$ scaling.
However, this faces with the notorious problem of measurement.

Hence, despite its wonderful mathematical structure and fast 
convergence, the CLB procedure is blocked off by the depth barrier. 
In the end, this due to the fact that then CLB matrix is nonlocal
hence it cannot be expressed as a compact product of the 2 qubit matrices 
of the native quantum gates. 

In light of the lesson learned with CLB, it is worth revisiting 
CNS in the hope of mitigating the depth problem.

\subsection{Carleman Navier-Stokes}

For the sake of simplicity let us consider the compressible Navier-Stokes 
equations 
\begin{eqnarray}\label{DNSE_cont}
\partial_t \rho + \partial_a J_a &=& 0\label{DNSE_rho_cont}\\
\partial_t J_a + \partial_b  \left[\frac{J_a J_b}{\rho} + P \delta_{ab} + \sigma_{ab}\right] &=& 0,\label{DNSE_J_cont}  
\end{eqnarray}
which, upon spatial discretization and advanced in time by means of a simple Forward-Euler scheme: 
\begin{eqnarray}\label{DNSE}
\rho(\vec x,t+\Delta t) &=& \rho - \Delta t D_a J_a\label{DNSE_rho}\\
J_a(\vec x,t+\Delta t) &=&  J_a  - \Delta t D_b  \left[\frac{J_a J_b}{\rho} + P \delta_{ab} + \sigma_{ab}\right], \label{DNSE_J}  
\end{eqnarray}

\noindent where the right hand side is evaluated at $(\vec x,t)$.
In the above, $\sigma_{ab}$ is the dissipative tensor and 
$D_a$ denotes the Centered Finite Difference operator described in 
section~\ref{app:locality}, which 
is the discretized version of the spatial derivative along $x_a$.
Latin indices $a,b$ run again over the space coordinates, repeated
indices being summed upon.

The discretization shown above, based on Forward-Euler time-stepping and Centered Finite Differences, may 
appear quite naive as it is well known to suffer from various forms of numerical instability; on the one hand, in the high Reynolds regime, such instabilities are usually cured by applying some form of \emph{upwinding} or by adding an artificial excess diffusion~\cite{LeVeque,Jaime-Peraire:qk}, on the other hand, at low Reynolds, issues related to local incompressibility regions in the flow warrant either the use of different representations for the dependent variables~\cite{ZIENKIEWICZ1990105,QUARTERONI2000505,brezzifortin} or the use of suitable projection techniques to filter out spurious modes~\cite{CHORIN}. Nonetheless, the simple discretization approach presented is attractive in the current context due to its \emph{locality} properties that are discussed below (see Appendix~\ref{app:locality}), and does therefore deserve to be tested and assessed.

The second level of the Carleman approximation requires an equation for 
the second order tensor $\rho^{-1} J_aJ_b$, which we approximate assuming weak compressibility of the fluid. 

Yet another problem would arise with a non-ideal equation of state $P=P(\rho)$
but for the present purposes we shall stipulate an linear ideal-gas relation
$P=\rho c_s^2$. 
With this position, we need equations for $J_{ab} \equiv J_a J_b$ and 
the cubic term $\rho J_{ab}$. The Euler forward method in Carleman form delivers the following set of equations written in matrix form as

\begin{eqnarray}\label{eq:CNS_1}
J_\alpha(\vec x,t+\Delta t) =&& A_{\alpha\beta}J_\beta(\vec x,t)+B_{\alpha\beta\gamma}J_{\beta\gamma}(\vec x,t)\nonumber\\
&&+C_{\alpha\beta\gamma\delta}J_{\beta\gamma\delta}(\vec x,t),
\end{eqnarray}

\noindent where the greek indices run from 0 to $d$ and we identified $J_0$ with $\rho$. The matrices $A,B$ and $C$ can be derived from Eqs.~\eqref{DNSE_rho} and \eqref{DNSE_J}, and they are made by combinations of the derivative operators $D_a$. Using the explicit form of the dissipative tensor for incompressible fluids, $\sigma_{ab}=\nu[\partial_a(J_b/\rho)+\partial_b(J_a/\rho)]$, their components are

\begin{eqnarray}
A_{00}&=&\mathds{1},\quad A_{0a}=-\Delta t D_a,\quad A_{a0}=-c_s^2\Delta t D_a\nonumber\\
A_{ab}&=&(\mathds{1}+2\nu \Delta t D^{(2)})\delta_{ab}+2\nu\Delta t  D_aD_b\nonumber\\
B_{ab0}&=&-\nu\Delta t  D^{(2)}\delta_{ab}-\nu\Delta t D_a D_b,\quad B_{abc}=-2\Delta t\delta_{ab}D_c\nonumber\\
C_{abc0} & = & \Delta t\delta_{ab}D_c,
\end{eqnarray}

\noindent and the other terms are 0.
By far, $A$ is the most populated matrix. In the case of a two-dimensional system its explicit form is

\begin{eqnarray}
A =&& \mathds{1}+\Delta t
\begin{pmatrix}
0 & -D_1 & -D_2\\
-c_s^2 D_1 & 2\nu  D^{(2)}& 0 \\
-c_s^2 D_2 & 0 &  2\nu  D^{(2)}
\end{pmatrix}\nonumber\\
&&+2\nu \Delta t\begin{pmatrix}
0 & 0 & 0 \\
0 & D_1D_1 & D_1D_2 \\
0 & D_2D_1 &  D_2D_2 
\end{pmatrix}.
\end{eqnarray}

\noindent  The tensor $B$ has components

\begin{eqnarray}
B_0&=&0,
B_1 = -\Delta t
\begin{pmatrix}
0 & 0 & 0\\
\nu  (D^{(2)}+D_1D_1) & 2D_1 & 2D_2\\
\nu  D_1D_2 & 0 &  0
\end{pmatrix},\nonumber\\
B_2 &=& -\Delta t
\begin{pmatrix}
0 & 0 & 0\\
\nu  D_2D_1 & 0 &  0\\
\nu  (D^{(2)}+D_2D_2) & 2D_1 & 2D_2
\end{pmatrix}
\end{eqnarray}

\noindent and the tensor $C_{[\dots]0}$ has non-null components

\begin{eqnarray}
C_{1[..]0} &=& \Delta t
\begin{pmatrix}
0 & 0 & 0\\
0 & D_1 & D_2\\
0 & 0 &  0
\end{pmatrix},
C_{2[..]0} = \Delta t
\begin{pmatrix}
0 & 0 & 0 \\
0 & 0 &  0\\
0 & D_1 & D_2 
\end{pmatrix}.
\end{eqnarray}

The evolution of the second order Carleman variables is obtained by multiplying the Eq.~\eqref{eq:CNS_1} by $J_\beta$, which delivers several terms of order two to six. 

The tensor product between the vectors $J_\alpha$ and $J_\beta$ leads to nonlocal terms with components $J_{\alpha\beta}(x,y)$ for any grid point $\vec x,\vec y$~\cite{CHILDS}, and with a total dimension $(n_vG)^2$. However, a much smaller vector can be obtained if we consider only the local variables $J_{\alpha\beta}(\vec x)\equiv J_{\alpha\beta}(\vec x,\vec x)$ which lead to a second order vector of dimension just $n_v^2G$.
Thus, the potential advantage of CNS over CLB relates to the count 
of Carleman variables, based not only on the number of variables 
per site but potentially also on better locality. In fact, the conservative form of Eqs.~\eqref{DNSE_rho},\eqref{DNSE_J} is preserved also at higher Carleman orders. All the terms can in fact be written under divergence and the derivative operator acts relating local variables at different locations. Nonetheless, the Carleman variables remain function of a single location. 
For instance, the discrete derivative applied to the first term in the square brackets in Eq.~\eqref{DNSE_J} in direction $\vec e_a$ assumes the form:

\begin{eqnarray}
D_aJ_{ab}(\vec x,t) = J_{ab}(\vec x+\vec e_a,t)-J_{ab}(\vec x,t)\\
D_aJ_{ab0}(\vec x,t) = J_{ab0}(\vec x+\vec e_a,t)-J_{ab0}(\vec x,t)
\end{eqnarray}

Note that in the LBM formulation, this role is taken by the streaming step, which requires the calculation of the non local variables $J_{ab}(\vec x,\vec y)$ at each grid point, see Fig.~\ref{fig:non_locality_CLB} and analogously for the higher order Carleman terms~\cite{SANAV}. This feature is further discussed in the Appendix.
 
At Carleman level $K=1$, we have $V_1 \equiv [\rho,J_a]$ 
which makes $4$ in $d=3$ against $27$ for LB, meaning two qubits for CNS and five for CLB.
With a $4^q$ depth, this means $4^2=16$ for CNS and $4^5 = 1024$ for CLB, which
is already a significant gain, even leaving nonlocality apart.

At level $K=2$, we have $V_2 = [\rho \rho, \rho J_a, J_a J_b]$, which makes 
another $1+3+6=10$, four qubits, against 
the $27 + 27 \times 28/2 =405$, $9$ qubits for LB. 
The gate count is $2^{8} = 256$ for CNS and
$2^{18} \sim 256K$ for CLB. The latter is too large for current
quantum computers but the former is viable.

\begin{figure*}
\includegraphics[scale = 0.52]{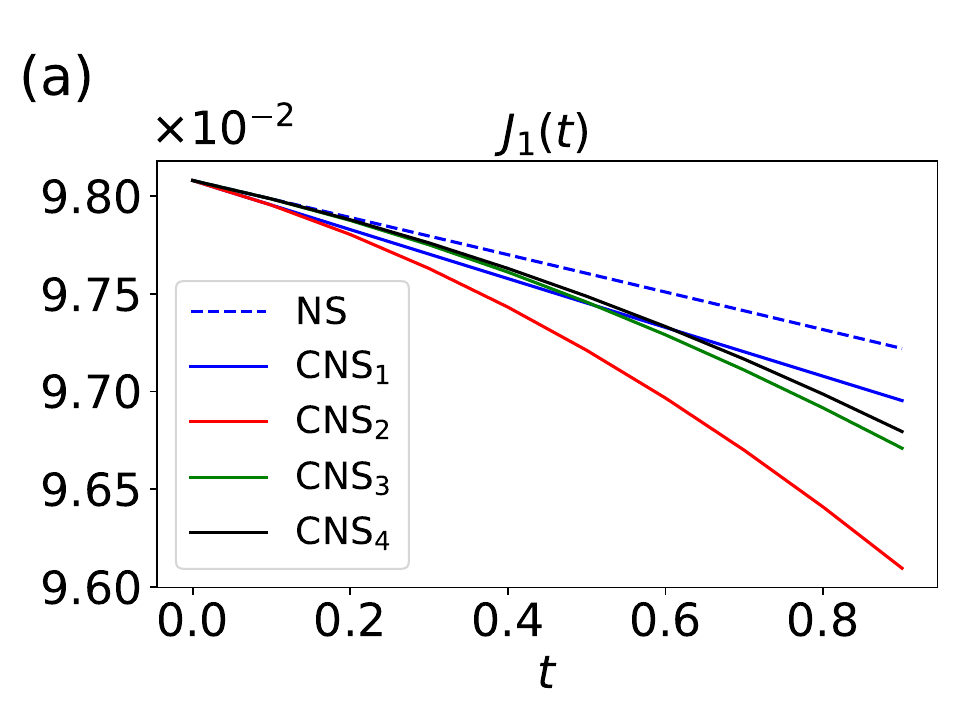}
\includegraphics[scale = 0.52]{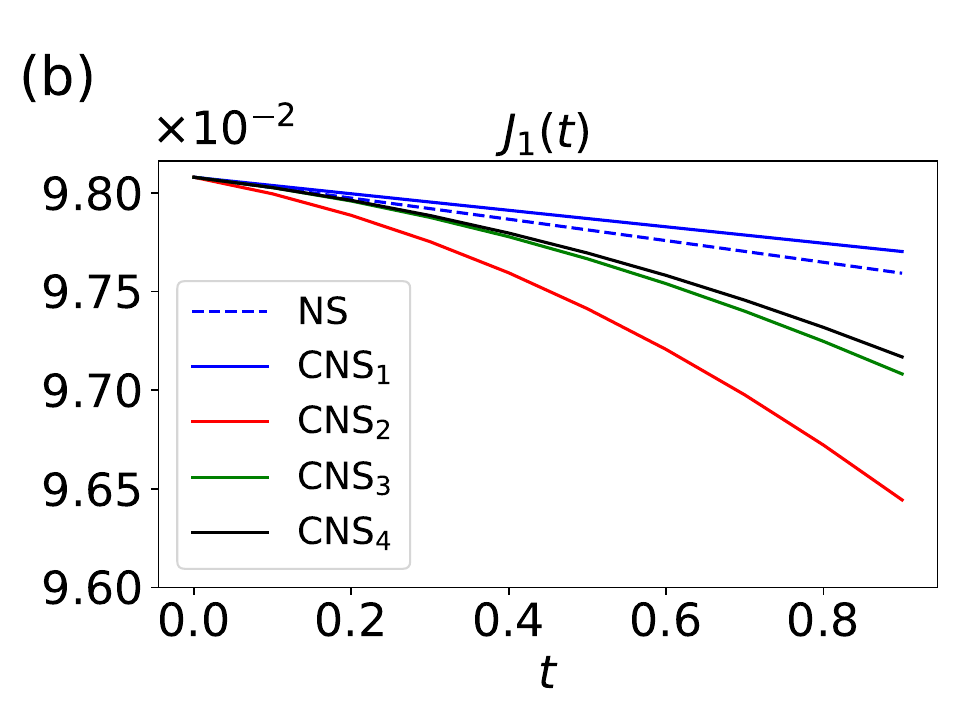}
\caption{\label{fig:CNS} The dynamics of the Kolmogorov-like flow on a $32\times32$ grid, plotted for a point of the grid, evolved with the Euler method from the NS equations~\eqref{DNSE_rho} and~\eqref{DNSE_J}, and comparison with CNS up to the fourth order for  $\nu =1/6$ in (a) and $\nu =1/18$ in (b). The parameters of the Kolmogorov flow at $t=0$ are $u_1 = u_2 = 0.1$, $k_1=k_2=1$.}
\end{figure*}

In order to make a fair comparison between the two methods, we analysed the Kolmogorov-like flow with the same initial conditions as defined in Eq.~\eqref{eq:kolmogorov-like}. We exploit Eq.~\eqref{eq:viscosity_omega} to derive the corresponding viscosity $\nu $, implicit in the NSE~\eqref{DNSE_J} in the stress-tensor term $\sigma_{ab}$.
We show in Fig.~\ref{fig:CNS} the solution obtained by Euler forward marching method for the same point plotted in Fig.~\ref{fig:CLB}, with $\nu =1/6$ in Fig.~\ref{fig:CNS}(a), corresponding to $\omega=1$, cf. \ref{fig:CLB}(a), and $\nu =1/18$ in Fig.~\ref{fig:CNS}(b), corresponding to $\omega=1.5$, cf. \ref{fig:CLB}(b).

From Figs.~\ref{fig:CNS}(a) and (b), we see that the convergence of CNS is valid only for small time frames up to O(10) timesteps. Increasing the Carleman order leads to a better approximation of the dynamics, even though the Carleman approximation remains valid for a very small time frame. 
We observe that the time frame of the CNS convergence is limited to less than O(100) time steps. We are aware 
of the fact that Euler time-marching with centered finite-differences is unconditionally unstable. However with time step currently used in the simulation ($\Delta t=0.01$), such instability 
is not expected to become apparent up to $t\sim 1$, since it is driven 
by a negative viscosity of order $\frac{U^2 \Delta t}{2} \simeq  5 \cdot 10^{-3}$. 
Furthermore, we wish to point out that in the time frame 
shown in Fig.\ref{fig:CNS} the density remains constant up to second digit. 
Finally, we are led to conclude that convergence of CNS is significantly poorer than CLB. 
This might be due to the Reynolds versus Mach argument.

Hence, contrary to what happens for CLB, CNS has convergence issues, which 
cannot be traced to the lack of stability of the Euler method. 
Much more work is needed to handle the behavior shown above. 
Incidentally, one may also consider resorting to different and possibly
more effcient linearization strategies than Carleman (see, {\it e.g},~\cite{QUARTERONI2000505})

\section{Carleman-Grad procedure}

Before closing, we discuss a third alternative which may combine the best
of CLB and CNS, namely the application of the Carleman procedure to
the Grad formulation of generalized hydrodynamics \cite{GRAD}.
The basic idea is to take progressive moments of the Boltzmann 
probability distribution and inspect the resulting open hierarchy
of hyperbolic PDE's. 
At order three  in the Grad expansion, we obtain
\begin{eqnarray}    
\label{eq:GRAD}
\partial_t \rho   + \partial_a J_a =0\\
\partial_t J_a    + \partial_b P_{ab} =0\\
\partial_t P_{ab} + \partial_c Q_{abc} = -\omega(P_{ab}-P_{ab}^{eq})
\end{eqnarray}    
where $\rho = \int f dv$ is the fluid density, $J_a = \int f v_a dv$
is the fluid current, $P_{ab} = \int f v_a v_b dv $ is the momentum flux
tensor and $Q_{abc} = \int f v_a v_b v_c dv$ is the energy flux tensor.

As is well known, this is a hyperbolic superset of the Navier-Stokes
equations, which are recovered in the limit of weak departure from local
equilibrium. Under such limit, the third equation can be closed, by
assuming adiabatic relaxation of $P_{ab}$ to its equilibrium, namely:
\begin{equation}
P_{ab} \sim P_{ab}^{eq} -\tau \partial_c Q_{abc}^{eq}
\end{equation}
where $\tau=1/\omega$ and the equilibrium expressions,
and
$Q_{abc}^{eq} = J_a J_b J_c/\rho^2 + J_a \delta_{bc} +J_b \delta_{ac} + J_c \delta_{ab}$.

The appeal of the Grad formulation is its hyperbolic character, which reflects
in the conservative nature of the equations.

Euler time marching delivers:
\begin{eqnarray}    
\label{GRADEUL}
\rho(t+\Delta t)   = \rho(t)   -\Delta tD_a J_a\\
J_a(t+\Delta t)    = J_a(t)    -\Delta tD_b P_{ab}\\
P_{ab}(t+\Delta t) = P_{ab}(t) -\Delta tD_c Q_{abc}^{eq} -\omega\Delta t(P_{ab}-P_{ab}^{eq})
\end{eqnarray}    
This scheme preserves locality because $P_{ab}$ is an independent 
variable under divergence. 

The first order Carleman step is to set the nonlinear terms
$J_{ab}=0$ and $J_{abc} \equiv J_a J_{bc}/\rho=0$.

\subsection{Picard iterations}

However, further Carleman steps require knowledge of $J_{ab}$ to 
compute $P_{ab}^{eq}$. The dynamic equation for $J_{ab}$ can be 
obtained the same way as for CNS, at the cost of introducing six additional
fields, for a total of $16$, namely still $4$ qubits. 

A possible alternative is to resort to Picard 
iteration by computing the above terms using the values of
$\rho$ and $J_a$ from the previous iteration.
In other words, by storing a separate copy of the Carleman 
variable at each iteration level $l$, one would compute 
$J_{ab}^{p+1} = (J_a^p J_b^{p+1} + J_a^{p+1} J_b^p)/(\rho^p+\rho^{p+1})$, which 
is linear in the Carleman variables $J_a^{p+1}$.
With $P$ Picard iterations this brings about another $4P$ 
Carleman variables. Assuming fast convergence, say $P=3$ takes 
the count to $10+12=22$, namely $5$ qubits per site and a 
circuit depth $2^{10}=1024$, possibly doable in the near future.  

Clearly, this iterative procedure could be applied to the CNS
framework as well, but the Grad picture offers a number of specific advantages.

First, like in LB, dissipation occurs via adiabatic relaxation 
of $P_{ab}$ to its equilibrium value, a fully local process which 
does not entail any communication in space, hence no 
Laplace operator as required in the NS picture. 
Second, again like in LB, the nonlinearity appears to be controlled by the Mach number.
Third, since $P_{ab}$ obeys its own conservative evolution equation, 
weak-compressibility effects are naturally incorporated with no need of ad-hoc numerics.  

As compared to CLB, the advantage is a lesser number of variables and
a non-directional streaming, which favors locality.
This said, the iterative procedure adds another layer of complexity, hence
its benefits must be carefully weighed against the corresponding costs.
Work is current underway to provide a quantitative assessment of the CLB
procedure.

\subsection{Summary of Carleman linearization}

Summarizing, it appears like Carleman linearization for fluids presents
a dual picture: on the one side, CLB is very compact and elegant,
but it involves an exponential growth of variable not only because it needs
many variables per site but because free-streaming couples variables across sites.

On the other side, the CNS framework offers a drastically reduced number
of Carleman variables, first because there are less variables per site and
potentially more locality, as long as  conservative difference schemes
can be developed.  

Finally, the CG framework may offer an optimal trade-off between the two.

Much further (hard) work is needed to assess whether any of the three
Carleman routes discussed in this paper may finally open the way
to the quantum simulation of classical fluids.

\section{Appendix: Local and non-local finite difference terms} 
\label{app:locality}

%
%
%
Using centered differences , with $\Delta x=\Delta y=1$, the 
divergence of a generic second order tensor $T_{ab}$ reads as follows:

\begin{eqnarray}
D_b T_{xb} =&& \frac{1}{2} [T_{xx}(i+1,j)-T_{xx}(i-1,j)+T_{xy}(i,j+1)\nonumber\\
&&-T_{xy}(i,j-1)]\\
D_b T_{yb} = &&\frac{1}{2} [T_{yx}(i+1,j)-T_{yx}(i-1,j)+T_{yy}(i,j+1)\nonumber\\
&&-T_{yy}(i,j-1)]       
\end{eqnarray}
These expressions are local since they do not involve any coupling between 
$T_{ab}$ at different sites, all we need to store is $T_{ab}(i,j)$.

Consider now a non-conservative second order term of the form:

\begin{eqnarray}
\rho D_a J_a = &&\rho(i,j) [\frac{1}{2} [J_x(i+1,j)-J_x(i-1,j) + J_y(i,j+1)\nonumber\\
&&-J_y(i,j-1)] 
\end{eqnarray}

Such term involves nonlocal products $\rho(i,j)J_x(i\pm 1,j)$ and
$\rho(i,j)J_y(i,j\pm 1)$, thus breaking the locality of 
the Carleman vector $V_a \equiv \rho J_a$. 
As a result we need to store $V_a(i,j;i \pm 1, j \pm 1))$, a non-local Carleman vector.
Next, write $\rho D_a J_a = D_a (\rho J_a) - J_a D_a \rho$.
In the limit of weak density gradients, the second term can be neglected
and the Carleman vector $V_a(i,j)$ is all we need to store: locality is resumed.   

\section*{Acknowledgements}

The authors have benefited from valuable discussions 
with many colleagues, particularly
S.S. Bharadwaj, D. Buaria,
P. Coveney, N. Defenu, 
G. Galli, M. Grossi, B. Huang, A. Mezzacapo, 
S. Ruffo, A. Solfanelli and T. Weaving.
S.S. and C.d.F. acknowledge financial support form the Italian National
Centre for HPC, Big Data and Quantum Computing (CN00000013). 

The authors have no conflicts to disclose.
The data that support the findings of this study are available from the corresponding author upon reasonable request.
\nocite{*}
\bibliography{aipsamp}

\bibliographystyle{aipauth4-2}

\end{document}